\documentclass[conference,a4paper]{IEEEtran}

\usepackage{amsfonts}
\usepackage{amssymb}
\usepackage{mathrsfs}
\usepackage{amsmath}
\usepackage{cite}
\usepackage{mathrsfs}
\usepackage[ruled,vlined]{algorithm2e}
\usepackage{tikz}
\usetikzlibrary{arrows,automata}
\usepackage[latin1]{inputenc}
\usepackage{verbatim}
\makeatletter

\newcommand{\Rmnum}[1]{\expandafter\@slowromancap\romannumeral #1@}
\makeatother

\newtheorem{thm}{Theorem}
\newtheorem{lemma}[thm]{Lemma}

\newtheorem{cor}[thm]{Corollary}
\newtheorem{defn}{Definition}
\newtheorem{rem}[thm]{Remark}

\newtheorem{rem-eg}[thm]{Remark and Example}

\newcommand{\ti}{\tilde}

\newcommand{\w}{{\omega}}
\newcommand{\p}{{\rho}}
\newcommand{\de}{{\delta}}
\newcommand{\dt}{{\delta}}

\newcommand{\bzero}{{\bf 0}}
\newcommand{\f}{\tilde{f}}
\newcommand{\g}{\tilde{g}}
\newcommand{\F}{\tilde{F}}
\newcommand{\mL}{\mathcal{L}}
\newcommand{\mF}{\mathcal{F}}
\newcommand{\mP}{\mathcal{P}}

\newcommand{\mE}{\mathcal{E}}
\newcommand{\mT}{\mathcal{T}}
\newcommand{\Rank}{{\mathrm{Rank}}}

\newcommand{\row}{{\rm row}}
\newcommand{\ul}{\underline}
\SetKw{KwInitialization}{Initialization:}

\begin{document}

\sloppy

%% Paper Title
%% You can use linebreaks \\ within to get better formatting as
%% desired.
\title{Linear Network Error Correction Multicast/Broadcast/Dispersion Codes}
%% Author names and affiliations:
%%
%% Avoiding spaces at the end of the author lines is not a problem with
%% conference papers because we don't use \thanks or \IEEEmembership.
%%
%% For several authors with only one affiliation:
%%
% \author{
%   \IEEEauthorblockN{Hui-Ting Chang and Stefan M.~Moser}
%   \IEEEauthorblockA{Department of Electrical and Computer Engineering\\
%     National Chiao Tung University (NCTU)\\
%     Hsinchu, Taiwan\\
%     Email: \{email-of-hui-ting,email-of-stefan\}@ieee.org}
% }
%%
%% For up to three affiliations:
%%
\author{
  \IEEEauthorblockN{Xuan~Guang}
  \IEEEauthorblockA{School of Mathematical Science and LPMC\\
    Nankai University\\
    Tianjin, P. R. China\\
    Email: xguang@nankai.edu.cn}
  \and
  \IEEEauthorblockN{Fang-Wei~Fu}
  \IEEEauthorblockA{Chern Institute of Mathematics and LPMC\\
    Nankai University\\
    Tianjin, P. R. China\\
    Email: fwfu@nankai.edu.cn}
}
%%
%% For over three affiliations, or if they all won't fit within the width
%% of the page, use this alternative format:
%%
% \author{
%   \IEEEauthorblockN{
%     Michael Shell\IEEEauthorrefmark{1},
%     Homer Simpson\IEEEauthorrefmark{2},
%     James Kirk\IEEEauthorrefmark{3},
%     Montgomery Scott\IEEEauthorrefmark{3} and
%     Eldon Tyrell\IEEEauthorrefmark{4}}
%   \IEEEauthorblockA{
%     \IEEEauthorrefmark{1}School of Electrical and Computer Engineering\\
%     Georgia Institute of Technology, Atlanta, Georgia 30332--0250\\
%     Email: see http://www.michaelshell.org/contact.html}
%   \IEEEauthorblockA{
%     \IEEEauthorrefmark{2}Twentieth Century Fox, Springfield, USA\\
%     Email: homer@thesimpsons.com}
%   \IEEEauthorblockA{
%     \IEEEauthorrefmark{3}Starfleet Academy, San Francisco, California 96678-2391\\
%     Telephone: (800) 555--1212, Fax: (888) 555--1212}
%   \IEEEauthorblockA{
%     \IEEEauthorrefmark{4}Tyrell Inc., 123 Replicant Street, Los Angeles, California 90210--4321}
% }
%% Use for special paper notices

%% To balance the two columns, you should reduce the text-height of
%% the last page using the following command:
%%%%%%%%%%%%%%%%%%%%%%%%%%%%%%%%%%%%%%%%%%%%%%%%%%%%%%%%%%%%%%%%%%%%%
%\addtolength{\textheight}{-9.35cm}
%%%%%%%%%%%%%%%%%%%%%%%%%%%%%%%%%%%%%%%%%%%%%%%%%%%%%%%%%%%%%%%%%%%%%
%% with an appropriate value. This command must be place on the second
%% last page, i.e., for a one-page abstract here, for a two-page
%% abstract right after the \maketitle command.
%% Create the title:
\maketitle
%% Abstract:
%% For the final version of the accepted paper, please make sure you
%% remove the comment "THIS PAPER IS ELIGIBLE FOR THE STUDENT PAPER
%% AWARD."
%%
\begin{abstract}
In this paper, for the purposes of information transmission and network error correction simultaneously, three classes of important linear network codes in network coding, linear multicast/broadcast/dispersion codes are generalized to linear network error correction coding, i.e., linear network error correction multicast/broadcast/dispersion codes. We further propose the (weakly, strongly) extended Singleton bounds for these new classes of codes, and define the optimal codes satisfying the corresponding Singleton bounds with equality, which are called multicast/broadcast/dispersion MDS codes respectively. The existence of such codes are  proved by an algebraic method and one kind of constructive algorithm is also proposed.
\end{abstract}

%%%%%%%%%%%%%%%%%%%%%%%%%%%%%%%%%%%%%%%%%%%%%%%%%%%%%%%%%%%%%%%%%%%%%%%%%%%%%%%%%%%%%%%%%%%%%%
\section{Introduction}
In network communication, the source node can multicast the information to all sink nodes at higher rate, if network coding is applied in a network, rather than routing alone, \cite{Ahlswede-Cai-Li-Yeung-2000}--\cite{Yeung-book}. When all kinds of errors may occur in network communication, network error correction coding (NEC) was discussed widely in order to deal with such problems, \cite{Yeung-Cai-correct-1}--\cite{Guang-MDS}. In this paper, three types of linear network error correction codes, linear network error correction multicast/broadcast/dispersion codes, are introduced and studied, which can be regarded as the generalization of three types of important linear network codes, linear network multicast/broadcast/dispersion codes.

A communication network is represented by a finite acyclic directed graph $G=(V,E)$, where $V$ and $E$ are the sets of nodes and channels of the network, respectively. A direct edge $e=(i,j)\in E$ stands for a channel leading from node $i$ to node $j$. Node $i$ is called the tail of $e$ and node $j$ is called the head of $e$, denoted by $tail(e)$ and $head(e)$, respectively. Correspondingly, the channel $e$ is called an outgoing channel of $i$ and an incoming channel of $j$. For a node $i$, define $Out(i)=\{e\in E:\ tail(e)=i\}$, and $In(i)=\{e \in E:\ head(e)=i\}$.
We allow the multiple channels between two nodes and assume that one field symbol can be transmitted over a channel in a unit time. In this paper, we only consider single source networks, and the unique source node is denoted by $s$.
Let $T$ be a collection of non-source nodes. A cut between the source node $s$ and $T$ is a set of channels whose removal disconnects $s$ from all $t\in T$. For unit capacity channels, the capacity of a cut between $s$ and $T$ can also be regarded as the number of channels in the cut, and the minimum of all capacities of cuts between $s$ and $T$ is called the minimum cut capacity $C_T$ between the source node $s$ and $T$. A cut between $s$ and $T$ of the minimum cut capacity is called a minimum cut. When the collection $T$ just consists of one node $t$, the above concepts degenerate into the general case for node $t$. In particular, the minimum cut capacity between the source node $s$ and non-source node $t$ is denoted as $C_t$. In fact, if we expand the single source network $G=(V,E)$ into $G_1=(V_1,E_1)$ by installing a new node $t_T$ which is connected from every node $t\in T$ by $|In(t)|$ multiple unit capacity channels, then the minimum cut capacity between the source node $s$ and $T$ in $G$ is equal to the minimum cut capacity between the source node $s$ and the node $t_T$ in $G_1$.

Let the information rate be $\w$ symbols per unit time. Then the source node has $\w$ imaginary incoming channels $In(s)=\{d_1',d_2',\cdots,d_\w'\}$. The source messages are $\w$ symbols $\underline{\bf{X}}=(X_1,X_2,\cdots,X_\w)\in\mF^{\w}$ arranged in a row vector.
When an error occurs on channel $e$, the output of the channel is $\tilde{U}_e=U_e+Z_e$, where $U_e$ is the message that should be transmitted over the channel $e$ and $Z_e\in \mF$ is the error occurred in $e$. In network $G=(V,E)$, corresponding to each channel $e\in E$, an imaginary channel $e'$ is introduced, which is connected to the tail of $e$ in order to provide error message $Z_e$. The network $\tilde{G}=(\tilde{V},\tilde{E})$ with imaginary channels is called the extended network of $G$, where $\tilde{V}=V$, $\tilde{E}=E\cup E'\cup In(s)$ with $E'=\{e': e\in E\}$. Obviously, $|E'|=|E|$. Then a linear network code on the original network $G$ can be amended to a linear network code on the extended network $\tilde{G}$ by setting local encoding coefficients $k_{e',e}=1$, $k_{e',d}=0$ for all $d\in E\backslash\{e\}$, and others $k_{d,e}\in \mF$ remains unchanged. Note that, for each non-source node $i$ in the extended network, $In(i)$ only includes the real incoming channels of $i$, that is, the imaginary channels $e'$ corresponding to $e\in Out(i)$ are not in $In(i)$. But for the source node $s$, we still define $In(s)=\{d_1',d_2',\cdots,d_\w'\}$. We can also define extended global encoding kernels $\f_e$ for all $e\in \tilde{E}$. The extended global encoding kernel $\f_e$ for $e\in \tilde{E}$ is an $(\w+|E|)$-dimensional column vector and the entries can be indexed by the elements of $In(s)\cup E$. For imaginary message channels $d_i'\ (1\leq i \leq \w)$ and imaginary error channels $e'\in E'$, let $\f_{d_i'}=1_{d_i'}$, $\f_{e'}=1_e$, where $1_d$ is an $(\w+|E|)$-dimensional column vector which is the indicator function of $d\in In(s)\cup E$. And for other global encoding kernels $\f_e, e\in E$, we have recursive formulae $\f_e=\sum_{d\in In(tail(e))}k_{d,e}\f_d+1_e$.

Let $\underline{\bf{Z}}=(Z_e:\ e\in E)$ be an $|E|$-dimensional row vector with $Z_e\in \mF$ for all $e\in E$, and $\underline{\bf{Z}}$ is called the error message vector. An error pattern $\p$ is regarded as a set of channels in which errors occur and we call that an error message vector $\underline{\bf{Z}}$ matches an error pattern $\p$, if $Z_e=0$ for all $e\in E\backslash \p$.

\begin{defn}
For an error pattern $\p$ and an extended global encoding kernel $\f_e$, $e\in E$, define:
\begin{itemize}
  \item $\f_e^{\p}$ an $(\w+|\p|)$-dimensional column vector obtained from $\f_e=(\f_e(d): d\in In(s)\cup E)$ by removing all entries $\f_e(d),d\notin In(s)\cup \p$.
  \item $f_e^{\p}$ an $(\w+|E|)$-dimensional column vector obtained from $\f_e=(\f_e(d): d\in In(s)\cup E)$ by replacing all entries $\f_e(d), d\notin In(s)\cup \p$ by $0$, and $f_e^{\p^c}\triangleq \f_e-f_e^{\p}$.
\end{itemize}
\end{defn}

%%%%%%%%%--Linear Network Error Correction Multicast/Broadcast/Dispersion MDS Codes--%%%%%%%%%%%%%%%%%
\section{Linear Network Error Correction Multicast/Broadcast/Dispersion Codes}
We introduce several new concepts. Some of which are regarded as the corresponding generalizations in original linear network error correction codes.

\begin{defn}
For a linear network error correction code on network $G$, let $T$ be a collection of non-source nodes and $In(T)=\cup_{t\in T}In(t)$. The decoding matrix $\ti{F}_T$ respect to $T$ is defined as:
$$\ti{F}_T=\begin{pmatrix}\f_e: e\in In(T) \end{pmatrix}=\begin{pmatrix}\f_e: e\in \cup_{t\in T}In(t) \end{pmatrix}.$$
\end{defn}
For the collection $T$, the decoding matrix $\F_T$ and $\ti{U}_e$, $e\in In(T)$ are available. And then, we can use the following equation
for decoding: $$\begin{pmatrix}\ul{{\bf X}}&\ul{{\bf Z}}\end{pmatrix}\cdot\F_T=\begin{pmatrix} \ti{U}_e: & e\in In(T) \end{pmatrix}.$$
Denote by $\row_T(d)$ the row vector of $\ti{F}_T$ indicated by the channel $d\in In(s)\cup E$.
Let $L$ be a collection of vectors in a vector space. We adopt the convention that $\langle L \rangle$ represents the subspace spanned by vectors in $L$. At a collection $T$ of non-source nodes, the following vector spaces are important.
\begin{defn}
For a linear network error correction code on network $G$ and a collection $T$ of non-source nodes, define
\begin{align*}
\Phi(T,G)&=\langle\{ \row_T(d_i'):\ i=1,2,\cdots,\w \}\rangle,\\
\Delta(T,\p,G)&=\langle\{ \row_T(e):\ e\in \p \}\rangle,
\end{align*}
which is called the message space of $T$ and the error space of error pattern $\p$ with respect to $T$, respectively.
\end{defn}
\begin{defn}
We say that an error pattern $\p_1$ is dominated by another error pattern $\p_2$ with respect to a collection $T$ of non-source nodes, if $\Delta(T,\p_1)\subseteq \Delta(T,\p_2)$ for any linear network error correction code. This relation is denoted by $\p_1 \prec_T \p_2$.
\end{defn}
\begin{defn}
The rank of an error pattern $\p$ with respect to a collection $T$ of non-source nodes is defined as
$$rank_T(\p)=\min\{ |\p'|:\ \p\prec_T \p'\}.$$
\end{defn}

In order to understand the above concept of rank of an error pattern, we give the following lemma.
\begin{lemma}\label{lem_rank}
Let $G=(V,E)$ be an acyclic network, $\p=\{e_1,e_2,\cdots,e_l\}$ be an error pattern on $G$ with $e_j\in In(i_j),\ (1\leq j \leq l)$, and $T$ be a collection of non-source nodes. Introduce a source node $s_\p$ and define $l$ channels $e_j'=(s_\p, i_j)$. Replace each $e_j$ by $e_j'$ $(1\leq j \leq l)$ on the network $G$, that is, add $e_1',e_2',\cdots,e_l'$ on the network and delete $e_1,e_2,\cdots,e_l$ from the network. Then the rank of the error pattern $\p$ with respect to the collection $T$ is equal to the minimum cut capacity between $s_\p$ and $T$.
\end{lemma}

The proof is similar to that of \cite[Lemma 1]{zhang-correction}, and, therefore, omitted. First, recall the concepts of linear multicast, linear broadcast, and linear dispersion, which have important applications in practice \cite{Li-Yeung-Cai-2003}\cite{Yeung-book}\cite{Tan-Yeung-Ho-Cai-Unified-Framework}.
\begin{defn}
An $\w$-dimensional linear network code on an acyclic network $G=(V,E)$ qualifies as a linear multicast, linear broadcast, and linear dispersion, respectively, if the following hold:
\begin{enumerate}
  \item $\dim(\mL(In(t)))=\w$ for every non-source node $t\in V$ with $C_t\geq \w$;
  \item $\dim(\mL(In(t)))=\min\{ \w, C_t \}$ for every non-source node $t\in V$;
  \item $\dim(\mL(In(T)))=\min\{ \w, C_T \}$ for every collection $T$ of non-source nodes;
\end{enumerate}
where $\mL(B)=\langle \{ f_e: e\in B \} \rangle$ for any subset $B\subseteq In(s)\cup E$.
\end{defn}

\begin{defn}\label{def_regular}
For an $\w$-dimensional linear network error correction code,
\begin{enumerate}
  \item it is called {\it regular}, if $\dim(\Phi(t))=\w$ for any non-source node $t\in V$ with $C_t\geq \w$.
  \item it is called {\it strongly regular}, if $\dim(\Phi(t))=\min\{\w, C_t\}$ for any non-source node $t\in V$.
  \item it is called {\it sup-regular}, if $\dim(\Phi(T))=\w$ for any collection $T$ of non-source nodes with $C_T\geq \w$.
  \item it is called {\it strongly sup-regular}, if $\dim(\Phi(T))=\min\{\w, C_T\}$ for any collection $T$ of non-source nodes.
\end{enumerate}
Further, a linear network error correction code is called multicast, broadcast and dispersion, if the corresponding regular, strongly regular and strongly sup-regular properties are satisfied respectively.
\end{defn}

%If a code is not regular, the code is not decodable at at least one sink node even in the error-free case. Therefore, we just consider the regular codes.
\begin{defn}
The minimum distance of a linear network error correction code on $G$ at any collection $T$ of non-source nodes is defined as
\begin{align*}
d_{\min}^{(T)}(G)&=\min\{ |\p|: \Delta(T,\p)\cap\Phi(T)\neq\{\ul{0}\} \}\\
&=\min\{ rank_T(\p): \Delta(T,\p)\cap\Phi(T)\neq\{\ul{0}\} \}\\
&=\min\{ \dim(\Delta(T,\p): \Delta(T,\p)\cap\Phi(T)\neq\{\ul{0}\} \}.
\end{align*}
Further, if $T$ is replaced by a non-source node $t\in V$, it is called the minimum distance of a linear network error correction code on $G$ at $t\in V$.
\end{defn}

Similar to original linear network error correction codes \cite{zhang-correction} \cite{Yang-refined-Singleton}\cite{Guang-MDS}, the above minimum distances fully characterize the error-detecting and error-correcting capabilities at the non-source node $t\in V$ and the collection $T$ of non-source nodes, respectively.

The same as in classical coding theory, some upper bounds on these minimum distances are of importance. The following upper bounds are similar to the Singleton bound in original linear network error correction codes \cite{Yeung-Cai-correct-1}\cite{zhang-correction}\cite{Yang-refined-Singleton}, and thus we call them the extended Singleton bounds.
\begin{thm}[Singleton Bounds]\label{thm_extended_Singleton}
Consider a strongly sup-regular linear network error correction code on a network $G=(V,E)$. Let $d_{\min}^{(T)}(G)$ be the minimum distance respect to a collection $T$ of non-source nodes in $V$. Then
\begin{align}\label{Singleton_T}
d_{\min}^{(T)}(G)\leq
\begin{cases}
C_T-\w+1 & \mbox{ if } C_T\geq \w,\\
1          & \mbox{ if } C_T < \w.
\end{cases}
\end{align}
This is called the extended Singleton bound. Similarly, replace the collection $T$ by one non-source node $t$, the above bound is called the weakly extended Singleton bound.
\end{thm}
\begin{rem}
Actually, it is easily seen that $d_{\min}^{(T)}(G)\geq 1$. Thus, for the case $C_T<\w$, we obtain $d_{\min}^{(T)}(G)=1$, which implies that there is no error-correcting capability. This is consistent with our intuition that the error correction is meaningless for the case $C_T<\w$, since none of source messages can be decoded at the collection $T$.
\end{rem}
\begin{IEEEproof}
Let $T$ be any collection of non-source nodes in $V$. Since the considered linear network error correction code is strongly sup-regular, it follows that
$\dim(\Phi(T))=\min\{\w, C_T\}$ from Definition \ref{def_regular}. To complete the proof, we discuss two cases below.

{\bf \textit{Case 1:}} $C_T\geq \w$, and thus $\dim(\Phi(T))=\w$.

Let the set of channels $\{e_1,e_2,\cdots,e_{C_T}\}$ be an arbitrary minimum cut between $s$ and $T$ with an upstream-to-downstream ancestral order $e_1\prec e_2\prec \cdots \prec e_{C_T}$, and $\p=\{e_{\w},e_{\w+1},\cdots,e_{C_T}\}$ be an error pattern. As we will show, $\Delta(T,\p)\cap\Phi(T)\neq \{\ul{0}\}$, which implies that $d_{\min}^{(T)}\leq C_T-\w+1$.

Let $\ul{X}$ be a source message vector and $\ul{Z}$ be an error message vector. Then for each channel $e\in E$, we know $\begin{pmatrix}\ul{X}&\ul{Z}\end{pmatrix}\cdot \ti{f}_e=\ti{U}_e$, where $\ti{f}_e$ is the extended global encoding kernel of $e$ and $\ti{U}_e$ is the output of the channel $e$. Since the rank of the matrix
$\begin{pmatrix}\ti{f}_{e_1}&\ti{f}_{e_2}&\cdots&\ti{f}_{e_{\w-1}} \end{pmatrix}$ is at most $(\w-1)$, there exists a nonzero message vector $\ul{X}_1$ and an all-zero error message vector $\ul{Z}_1=\ul{0}$ such that
\begin{align*}
(\ul{X}_1\ \ul{Z}_1)\cdot(\ti{f}_{e_1}\ \ti{f}_{e_2}\ \cdots\ \ti{f}_{e_{\w-1}})=(\ti{U}_{e_1}\ \ti{U}_{e_2}\ \cdots\ \ti{U}_{e_{\w-1}} )=\ul{0}.
\end{align*}
Furthermore, since the linear network error correction code is strongly sup-regular, this further implies
\begin{align*}
(\ul{X}_1\ \ul{Z}_1)\cdot(\ti{f}_{e_1}\ \ti{f}_{e_2}\ \cdots\ \ti{f}_{e_{C_T}}
)=(\ti{U}_{e_1}\ \ti{U}_{e_2}\ \cdots\ \ti{U}_{e_{C_T}})\neq\ul{0}.
\end{align*}
Assume the contrary that $(\tilde{U}_{e_1}\ \tilde{U}_{e_2}\ \cdots\ \tilde{U}_{e_{C_T}})=\ul{0}$, which further implies
$\tilde{A}_t\triangleq(\tilde{U}_e: e\in In(T))=\underline{0}$, since $\{e_1,e_2,\cdots,e_{C_T}\}$ is a minimum cut between $s$ and $T$ and $\ul{Z}_1=\ul{0}$. Furthermore, we have $(\underline{X}_1\ \underline{0})\tilde{F}_T=\underline{0}$ from the decoding equation
$(\underline{X}_1\ \underline{Z}_1)\tilde{F}_T=\tilde{A}_T$. Therefore, $\underline{X}_1=\underline{0}$ from $\dim(\Phi(T))=\w$. This contradicts to our assumption $\underline{X}_1\neq\underline{0}$.

On the other hand, there exists another source message vector $\ul{X}_2=\ul{0}$ and another error message vector $\ul{Z}_2$ matching the error pattern $\p=\{e_{\w},e_{\w+1},\cdots,e_{C_T}\}$, such that
$$
\begin{pmatrix}\ul{X}_2&\ul{Z}_2\end{pmatrix}\cdot\begin{pmatrix}\ti{f}_{e_1}&\cdots&\ti{f}_{e_{C_T}} \end{pmatrix}
=\begin{pmatrix}\ti{U}_{e_1}&\cdots&\ti{U}_{e_{C_T}} \end{pmatrix}.
$$
Note that $\underline{Z}_2\neq \underline{0}$ because $(\tilde{U}_{e_1}\ \tilde{U}_{e_2}\ \cdots\ \tilde{U}_{e_{C_T}})\neq \underline{0}$.
Since $e_{\w}\prec e_{\w+1}\prec \cdots\prec e_{C_T}$, for any $e\in \p$, we can set sequentially with the boundary condition $Z_d=0$ for all $d\in E\backslash\p$:
$$Z_e=\ti{U}_e-\sum_{d\in In(tail(e))}k_{d,e}\ti{U}_d',$$
where $\tilde{U}_d'$ is the output of channel $d$ in this case.

Combining the above, we have
$$\begin{pmatrix}\ul{X}_1&\ul{0}\end{pmatrix}\cdot\ti{F}_T
=\begin{pmatrix}\ul{0}&\ul{Z}_2\end{pmatrix}\cdot\ti{F}_T,$$
which, together with the fact that $\ul{Z}_2\neq \ul{0}$ matches the error pattern $\p$, proves that
$$\Phi(T)\cap\Delta(T,\p)\neq \{\ul{0}\}.$$
That is, $d_{\min}^{(T)}(G)\leq C_T-\w+1$ for any collection of non-source nodes $T$ with $C_T\geq \w$.

{\bf \textit{Case 2:}} $C_T<\w$ and thus $\dim(\Phi(T))=C_T<\w$. Similarly, apply the same method to a minimum cut $\{e_1,e_2,\cdots,e_{C_T}\}$ and an error pattern $\p=\{e_{C_T}\}$. We can obtain $d_{\min}^{(T)}(G)\leq 1$ for all collections $T$ of non-source nodes with $C_T<\w$. Combining the two cases, the proof is completed.
\end{IEEEproof}

Adopt the convention that the codes satisfying Singleton bound with equality are called maximum distance separable (MDS) codes.
\begin{defn}
An $\w$-dimensional linear network error correction code over a network $G$ is called linear network error correction multicast/broadcast/dispersion MDS code, respectively, or multicast/broadcast/dispersion MDS code for short , if the following hold:
\begin{enumerate}
  \item it is regular and $d_{\min}^{(t)}(G)=C_t-\w+1$ for any non-source node $t\in V$ with $C_t\geq \w$;
  \item it is strongly regular and the weakly extended Singleton bound is satisfied with equality for any non-source node $t\in V$;
  \item it is strongly sup-regular and the extended Singleton bound (\ref{Singleton_T}) is satisfied with equality for any collection $T$ of non-source nodes.
\end{enumerate}
\end{defn}
%\begin{rem}\label{rem_relation-m-b-d}
%It is not difficult to observe that dispersion MDS codes are broadcast MDS codes, and broadcast MDS codes are multicast MDS codes,
%and the broadcast MDS codes are the same as the multicast MDS codes, except the sup-regular property.
%\end{rem}

%%%%%%%%%%%%%%%%%%%%%%%%%%%%%%%%%%%%%%%%%%%%%%%%%%%%%%%%%%%%%%%%%%%%%%%%%%%%%%%%%%%%%%%%

\section{The Existence of the Optimal Codes}

In this section, we show that multicast/broadcast/dispersion MDS codes defined above do exist which means that our proposed Singleton bounds are achievable. For a collection $T$ of non-source nodes with $C_T\geq \w$ and a non-source node $t\in V$ with $C_t\geq \w$, define:
\begin{align*}
R_T(\dt_T,G)&=\{\mbox{ error pattern }\p:\ |\p|=rank_T(\p)=\dt_T \},\\
R_t(\dt_t,G)&=\{\mbox{ error pattern }\p:\ |\p|=rank_t(\p)=\dt_t\},
\end{align*}
where $\dt_T=C_T-\w+1$ and $\dt_t=C_t-\w+1$. When there is no ambiguity, $R_T(\dt_T,G)$ and $R_t(\dt_t,G)$ will be abbreviated as $R_T(\dt_T)$ and $R_t(\dt_t)$, respectively.
\begin{thm}\label{thm_e_m}
For a network $G=(V,E)$, there exists an $\w$-dimensional linear network error correction multicast MDS code on $G$, if the size of the base field satisfies:
$$|\mF|>\sum_{t\in V: C_t\geq \w}|R_t(\dt_t)|.$$
\end{thm}

In order to prove this result, we need to prepare several lemmas below.
\begin{lemma}[{\cite[Corollary 4]{Guang-MDS}}]\label{lem_path}
For each $t\in V$ with $C_t\geq \w$ and each error pattern $\p\in R_t(\dt_t)$, there exist $(\w+\dt_t)$ channel disjoint paths from either $In(s)=\{d_1',d_2',\cdots,d_\w'\}$ or $\p'=\{e': e\in \p\}$ to $t$ satisfying the following properties:
\begin{enumerate}
  \item there are exactly $\dt_t$ paths from $\p'$ to $t$, and $\w$ paths from $In(s)$ to $t$;
  \item these $\dt_t$ paths from $\p'$ to $t$ start with the distinct channels in $\p'$ and for each path, if it starts with $e'\in \p'$, then it passes through $e\in \p$.
\end{enumerate}
\end{lemma}
\begin{lemma}[{\cite[Lemma 1]{Koetter-Medard-algebraic}}]\label{lem_poly}
Let $f(x_1,x_2,\cdots,x_n)$ be a nonzero polynomial with coefficients in a field $\mF$. If $|\mF|$ is greater than the degree of $f$ for any $x_i$ $(1\leq i \leq n)$, then there exist $a_1,a_2,\cdots,a_n\in \mF$ such that
$f(a_1,a_2,\cdots,a_n)\neq 0$.
\end{lemma}
\begin{lemma}[{\cite[Lemma 2]{zhang-correction}}]\label{lem_min_dis}
A code is regular and has minimum distance $d_{\min}^{(t)}\geq d+1$ if
$$rank_t(\p)\geq d\ \Longrightarrow\ \Rank(\ti{F}_t^{\p})\geq \w+d.$$
\end{lemma}
\begin{IEEEproof}[Proof of Theorem \ref{thm_e_m}]
Let $t$ be an arbitrary non-source node in $V$ with $C_t\geq \w$, and $\p$ be an arbitrary error pattern in $R_t(\de_t)$. It is not difficult to see that each entry of the decoding matrix $\ti{F}_t$ (obviously, $\ti{F}_t^{\p}$) is a polynomial of local encoding coefficients $k_{d,e}$ for some channel adjacent pairs $(d,e),\ d,e\in In(s)\cup E$. For the non-source node $t$ and the error pattern $\p$, there exist $(\w+\de_t)$ channel disjoint paths satisfying the following conditions from $rank_t(\p)=\de_t$ and Lemma \ref{lem_path}: 1) there are exactly $\de_t$ paths from $\p'$ to $t$, and $\w$ paths from $In(s)$ to $t$; 2) these $\de_t$ paths from $\p'$ to $t$ start with the distinct channels in $\p'$ and for each path, if it starts with $e'\in \p'$, then it passes through $e\in \p$.
Further, let $In'(t)$ be the collection of the last channels of the $\w+\de_t$ paths. Clearly, $In'(t)\subseteq In(t)$. For any subset $\eta$ of $E$, define a $|\eta|\times|E|$ matrix $A_{\eta}=(A_{d,e})_{d\in \eta, e\in E}$ satisfying:
$$A_{d,e}=\begin{cases} 1 & d=e;\\
                        0 & \text{otherwise.}\end{cases}$$
Recall the matrix $\ti{M}=\begin{pmatrix}\ti{f}_e: e\in E\end{pmatrix}$, where all extended global encoding kernels are put in juxtaposition according to the ancestral order, the matrix $\ti{B}=\left(\begin{smallmatrix} B \\ I\end{smallmatrix}\right)$, where $I$ denotes an $|E|\times|E|$ identity matrix and $B=(k_{d,e})_{d\in In(s), e\in E}$ is an $\w\times|E|$ matrix with $k_{d,e}=0$ for $e\notin Out(s)$ and $k_{d,e}$ being the local encoding coefficient for $e\in Out(s)$, and the system transfer matrix $K=(k_{d,e})_{d\in E, e\in E}$ is an $|E|\times|E|$ matrix where $k_{d,e}$ is the local encoding coefficient for $head(d)=tail(e)$ and $k_{d,e}=0$ for $head(d)\neq tail(e)$. Then we have formula $\ti{M}=\ti{B}(I-K)^{-1}.$
This formula (similar to the Koetter-M\'{e}dard Formula\cite{Koetter-Medard-algebraic}) first appears in \cite{zhang-correction}.
Define $\ti{A}_{\p}=\begin{pmatrix} I_{\w\times \w} & \bzero_{\w\times |E|}\\\bzero_{|\p|\times |E|} & A_\p \end{pmatrix}$ a matrix of size $(\w+|\p|)\times (\w+|E|)=(\w+\de_t)\times (\w+|E|)$, where $\bzero_{a\times b}$ represents an $a\times b$ zero matrix. Thus, we have
\begin{align*}
\ti{F}_t^{\p}\triangleq \ti{A}_{\p}\ti{B}(I-K)^{-1}A_{In(t)}^\top=\begin{pmatrix} B(I-K)^{-1}A_{In(t)}^\top\\A_{\p}(I-K)^{-1}A_{In(t)}^\top \end{pmatrix}.
\end{align*}
$$(\ti{F}_t^{\p})_{In'(t)}=\ti{A}_{\p}\ti{B}(I-K)^{-1}A_{In'(t)}^\top=\begin{pmatrix} B(I-K)^{-1}A_{In'(t)}^\top\\A_{\p}(I-K)^{-1}A_{In'(t)}^\top \end{pmatrix}$$
is a submatrix of $\ti{F}_t^{\p}$ and just consists of the column vectors of $\ti{F}_t^{\p}$ corresponding to the channels in $In'(t)$.
Thus put $k_{d,e}=1$ for all adjacent pairs of channels $(d,e)$ along every one of the chosen $(\w+\de_t)$ channel disjoint paths, and $k_{d,e}=0$, otherwise. It follows that $(\ti{F}_t^{\p})_{In'(t)}=I_{\w+\de_t}$, an $(\w+\de_t)\times (\w+\de_t)$ identity matrix, that is, $\det((\ti{F}_t^{\p})_{In'(t)})=1$,
which means that $\det((\ti{F}_t^{\p})_{In'(t)})$ is a nonzero polynomial.

Next, we show that the degree of each indeterminate $k_{d,e}$ in nonzero polynomial $\det((\ti{F}_t^{\p})_{In'(t)})$ is $1$ at most. Notice that
\begin{equation*}
\begin{split}
&\det( \begin{pmatrix}\ti{A}_{\p}\ti{B}&\bzero_{(\w+\de_t)\times(\w+\de_t)}\\I-K& A_{In'(t)}^\top \end{pmatrix} )\\
=&\det( \begin{pmatrix}\ti{A}_{\p}\ti{B}&-\ti{A}_{\p}\ti{B}(I-K)^{-1}A_{In'(t)}^\top\\I-K&\bzero_{\mE\times (\w+\de_t)}\end{pmatrix} )\\
=&(-1)^*\det((I-K))\cdot\det(\ti{A}_{\p}\ti{B}(I-K)^{-1}A_{In'(t)}^\top)\\
=&\det((\ti{F}_t^{\p})_{In'(t)})\cdot(-1)^*,
\end{split}
\end{equation*}
where $\det((I-K))=1$ because $K$ is an upper triangular matrix and all elements of main diagonal are zeros. This implies that the degree of each indeterminate $k_{d,e}$ in the polynomial $\det((\ti{F}_t^{\p})_{In'(t)})$ is $1$ at most. Further, applying it to every non-source node $t\in V$ with $C_t\geq \w$ and every $\p\in R_t(\de_t)$, it follows that
$$\prod_{t\in V: C_t\geq \w}\prod_{\p\in R_t(\de_t)}\det((\ti{F}_t^{\p})_{In'(t)})$$
is also a nonzero polynomial over the base field $\mF$, and the degree of each indeterminate $k_{d,e}$ is at most $\sum_{t\in V:C_t\geq \w}|R_t(\de_t)|$. Together with Lemma \ref{lem_poly}, this proves that if
$$|\mF|>\sum_{t\in V:C_t\geq \w}|R_t(\de_t)|,$$
we can choose scalar values in $\mF$ for all indeterminates such that
$$\prod_{t\in V: C_t\geq \w}\prod_{\p\in R_t(\de_t)}\det((\ti{F}_t^{\p})_{In'(t)})\neq 0,$$
which means all determinants are nonzero, that is, $\Rank((\ti{F}_t^{\p})_{In'(t)})=\w+\de_t$ for all $t\in T$ and $\p\in R_t(\de_t)$.

As $(\ti{F}_t^{\p})_{In'(t)}$ is a submatrix of $\ti{F}_t^{\p}$ with the same number of rows which is equal to the rank of $(\ti{F}_t^{\p})_{In'(t)}$, it is shown that $\Rank(\ti{F}_t^{\p})=\w+\de_t$. In addition, wherever $\p\prec_t \eta$, $\Rank(\ti{F}_t^{\p})\leq \Rank(\ti{F}_t^{\eta})$, together with Lemma \ref{lem_min_dis}, implies that
$d_{\min}^{(t)}(G)\geq \de_t+1$
for all non-source nodes $t\in V$ with $C_t\geq \w$. The proof is completed by combining what we have proved and weakly extended Singleton bound.
\end{IEEEproof}

Moreover, the existence of multicast MDS codes leads to the existence of broadcast/dispersion MDS codes.
\begin{thm}\label{thm_e_b}
There exist $\w$-dimensional linear network error correction broadcast/dispersion MDS codes on a network $G$, if the size of the base field satisfies respectively:
\begin{align*}
|\mF|&>\sum_{t\in V: C_t\geq \w}|R_t(\dt_t)|+|V_2|,\\
|\mF|&>\sum_{T\in \mT: C_T\geq \w}{|E|+\sum_{T\in \mT}\sum_{t\in T}C_t\choose \dt_T}+|V_3|,
\end{align*}
where $V_2\subseteq V$ is the set of all non-source nodes $t\in V$ satisfying $C_t<\w$, $\mT$ is the set of all collections of non-source nodes, and $V_3$ is the set of all collections $T\in \mT$ satisfying $C_T<\w$.
\end{thm}

We give the following corollary with the looser lower bound on the size of the base field $\mF$.
\begin{cor}
There exists an $\w$-dimensional linear network error correction multicast/broadcast MDS code on a network $G$, if the size of the base field satisfies respectively:
$$|\mF|>\sum_{t\in V:\ C_t\geq \w}{|E|\choose \de_t},
\mbox{ and }
|\mF|>\sum_{t\in V:\ C_t\geq \w}{|E|\choose \de_t}+|V_2|.$$
%where $V_2\subseteq V$ is the set of all non-source nodes $t\in V$ satisfying $C_t<\w$.
\end{cor}

%%%%%%%%%%%%%%%%%%%%%%%%%%%%%%%%%%%%%%%%%%%%%%%%%%%%%%%%%%%%%%%%%%%%%%%%%%%%%%%%%%%%%%%%%%%%%%%%%%%%%%%%%%%
\section{Construction of the Optimal Codes}

Actually, if we can design an algorithm for constructing multicast MDS codes, then it is not difficult to obtain algorithms for constructing broadcast/dispersion MDS codes. In the following, by modifying the algorithm for constructing network MDS codes proposed by Guang {\it et al.} \cite[Algorithm 1]{Guang-MDS}, we give an algorithm below for constructing multicast MDS codes.
First, we introduce some notation.
For arbitrary subset $B\subseteq In(s)\cup E\cup E'$, define
$\tilde{\mL}(B)=\langle \{ \f_e: e\in B \} \rangle$, $\tilde{\mL}^{\p}(B)=\langle \{ \f^{\p}_e: e\in B \} \rangle$, $\mL^{\p}(B)=\langle \{ f^{\p}_e: e\in B \} \rangle$, and $\mL^{\p^c}(B)=\langle \{ f^{\p^c}_e: e\in B \} \rangle.$
\begin{algorithm}[h]\label{algo_m}
%\dontprintsemicolon
\SetAlgoLined
%\SetLine
\KwIn{The network $G=(V,E)$, and the information rate $\w$.}
\KwOut{Extended global kernels (forming a multicast MDS code).}
\KwInitialization{
\rm For each non-source node $t\in V$ with $C_t\geq \w$ and each $\p\in R_t(\dt_t)$, choose $(\w+\dt_t)$ channel disjoint paths from $In(s)$ or $\p'$ to $t$ satisfying Lemma {\ref{lem_path}}. Denote by $\mP_{t,\p}$ the set of the above $(\w+\dt_t)$ channel disjoint paths, and $E_{t,\p}$ denotes the set of all channels on paths in $\mathcal{P}_{t,\p}$. Initialize dynamic channel set $CUT_{t,\p}=In(s)\cup\p'=\{d_1',\cdots,d_\w'\}\cup\{e': e\in\p\}$, as well as the extended global encoding kernels $\f_e=1_e$ for all imaginary channels $e\in In(s)\cup E'$.
}

\ForEach{\rm node $i\in V$ {\rm(according to the ancestral order)}}{
  \ForEach{\rm channel $e\in Out(i)$ }{
    \eIf{$e\notin \cup_{t\in V: C_t\geq \w}\cup_{\p\in R_t(\dt_t)}E_{t,\p}$,}{
       $ \tilde{f}_e=1_e$, all $CUT_{t,\p}$ remain unchanged.}{
    choose $\g_e\in\tilde{\mL}(In(i)\cup\{e'\})\backslash\cup_{t\in V:\atop C_t\geq \w}\cup_{\p\in R_t(\dt_t):\atop e\in E_{t,\p}}[\mL^{\p}(CUT_{t,\p}\backslash \{e(t,\p)\})+\mL^{\p^c}(In(i)\cup \{e'\})]$, where $e(t,\p)$ represents the previous channel of $e$ on the path which $e$ locates on.\\
       \eIf{$\g_e(e)=0$}{
          $\f_e=\g_e+1_e$\;}{
       $\f_e=\g_e(e)^{-1}\cdot\g_e$.}
       For those $CUT_{t,\p}$ satisfying $e\in E_{t,\p}$, update $CUT_{t,\p}=\{CUT_{t,\p}\backslash\{e(t,\p)\}\}\cup\{e\}$\; for others, $CUT_{t,\p}$ remain unchanged.
       }
}
}
\caption{Constructing a multicast MDS code.}
\end{algorithm}
\begin{rem}
Similarly, we  also can extend another important linear network code---generic code to network error correction. Because of the limit of pages, we omit that part. Cai \cite{Cai} proposed strongly generic linear network codes and very briefly discussed its application for network error correction. But this strongly generic condition is too strong for this application. Actually, the better conclusions can be obtained.
\end{rem}

%%%%%%%%%%%%%%%%%%%%%%%%%%%%%%%%------Acknowledgment------%%%%%%%%%%%%%%%%%%%%%%%%%%%%%%%%%%%
\section*{Acknowledgment}
The authors would like to thank Prof. Z. Zhang for his comments. This research is supported by the National Key Basic Research Program of China (973 Program Grant No. 2013CB834204), the National Natural Science Foundation of China (Nos. 61171082, 60872025, 10990011), and Fundamental Research Funds for the Central Universities of China (No. 65121007).

%%%%%%%%%%%%%%%%%%%%%%%%%%%%%%%%%%%%%%%%%%%%%%%%%%%%%%%%%%%%%%%%%%%%%%%%%%%%


\begin{thebibliography}{1}
%\bibitem{Zhang-Yeung-1999}
%R. W. Yeung and Z. Zhang, ``Distributed source coding for satellite communications,''
%\textit{IEEE Trans. Inf. Theory}, vol. 45, no. 4, pp. 1111-1120, May 1999.

\bibitem{Ahlswede-Cai-Li-Yeung-2000}
R. Ahlswede, N. Cai, S.-Y. R. Li, and R. W. Yeung, ``Network information
flow,'' \textit{IEEE Trans. Inf. Theory}, vol. 46, no. 4, pp. 1204-1216, Jul. 2000.

\bibitem{Li-Yeung-Cai-2003}
S.-Y. R. Li, R. W. Yeung, and N. Cai, ``Linear network coding,'' \textit{IEEE
Trans. Inf. Theory}, vol. 49, no. 2, pp. 371-381, Jul. 2003.

\bibitem{Koetter-Medard-algebraic}
R. Koetter and M. M$\acute{\textup{e}}$dard, ``An algebraic approach
to network coding,'' \textit{IEEE/ACM Trans.  Netw.}, vol. 11, no. 5,
pp. 782-795, Oct. 2003.

\bibitem{co-construction}
S. Jaggi, P. Sanders, P. A. Chou, M. Effros, S. Egner, K. Jain, and
L. M. G. M. Tolhuizen, ``Polynomial time algorithms for multicast
network code construction,'' \textit{IEEE Trans. Inf. Theory}, vol. 51, no. 6,
pp. 1973-1982, Jun. 2005.

\bibitem{Zhang-book}
R. W. Yeung, S.-Y. R. Li, N. Cai, and Z. Zhang, ``Network coding theory,''
\textit{Foundations and Trends in Communications and Information Theory}, vol. 2, nos.4 and 5, pp. 241-381, 2005.

\bibitem{Yeung-book}
R. W. Yeung, \textit{Information Theory and Network Coding}. Springer, 2008.

%\bibitem{Yeung-Cai-coorrect}
%N. Cai and R. W. Yeung, ``Network coding and error correction,'' \textit{in
%Proc. IEEE Inf. Theory Workshop}, Bangalore, India, Oct. 2002, pp. 119-122.

\bibitem{Yeung-Cai-correct-1}
R. W. Yeung and N. Cai, ``Network error correction, part \Rmnum{1}: Basic
concepts and upper bounds,'' \textit{Commun. Inf. Syst.}, vol. 6, pp. 19-36, 2006.

\bibitem{Yeung-Cai-correct-2}
N. Cai and R. W. Yeung, ``Network error correction, part \Rmnum{2}: Lower
bounds,'' \textit{Commun. Inf. Syst.}, vol. 6, pp. 37-54, 2006.

\bibitem{zhang-correction}
Z. Zhang, ``Linear network error correction codes in packet networks,'' \textit{IEEE Trans. Inf. Theory}, vol. 54, no. 1, pp. 209-218, Jan. 2008.

\bibitem{Yang-refined-Singleton}
S. Yang, R. W. Yeung, C. K. Ngai, ``Refined coding bounds and code constructions for coherent network error correction,'' \textit{IEEE Trans. Inf. Theory}, vol. 57, no. 3, pp. 1409-1424, Mar. 2011.

\bibitem{Guang-MDS}
X. Guang, F.-W. Fu, and Z. Zhang, ``Construction of network error correction codes in packet networks,''
\textit{IEEE Trans. Inf. Theory}, vol. 59, no. 2, pp. 1030-1047, Feb. 2013.

\bibitem{Tan-Yeung-Ho-Cai-Unified-Framework}
M. Tan, R. W. Yeung, S.-T. Ho, and N. Cai, ``A unified framework for linear network coding,''
\textit{IEEE Trans. Inf. Theory}, vol. 57, no. 1, pp. 416-423, Jan. 2011.

\bibitem{Cai}
N. Cai ``Valuable messages and random outputs of channels in
linear network coding,''  in \textit{Proc. IEEE Int. Symp. Inf. Theory}, Seoul, Korea, Jun. 2009, pp. 413--417.




%\bibitem{Yeung-relation-dispersion-generic}
%P.-W. Kwok and R. W. Yeung, ``On the relation between linear dispersion and generic network code,''
%{\it in proc. IEEE Information Theory Workshop 2006}, Chengdu, China, Oct. 2006, pp.413-417.

%\bibitem{Yeung-Framework}
%M. Tan, R. W. Yeung, and S. T. Ho, ``A unified framework for linear network codes,''
%{\it in the 4th Workshop on Network Coding, Theory and Applications
%(NetCod 2008)}, Hong Kong, China, Jan. 2008.

%\bibitem{Guang-uni-MDS}
%X. Guang, F.-W. Fu, and Z. Zhang, ``Universal Network Error Correction MDS Codes,'' to appear in NetCod 2011.

\end{thebibliography}
\end{document}